\documentclass[aps,prb,twocolumn,superscriptaddress]{revtex4}
\usepackage{bm}
\usepackage{graphicx,amsmath,latexsym,amssymb}
\usepackage{dcolumn}
\usepackage{hyperref}

\usepackage[usenames]{color}
\begin{document}

\definecolor{olive}{rgb}{0,1,.4}
\title{Localization of Ultrasound in a Three-Dimensional Elastic Network\footnote{Paper published in {\em Nature Physics} {\bf 4,} 945--948 (2008).\\
DOI:10.1038/nphys1101.\\
URL of the published article: \url{http://dx.doi.org/10.1038/nphys1101}.}}

\author{H. Hu}
\affiliation{Department of Physics and Astronomy, University of
Manitoba, Winnipeg, Manitoba, R3T 2N2 Canada}

\author{A. Strybulevych}
\affiliation{Department of Physics and Astronomy, University of
Manitoba, Winnipeg, Manitoba, R3T 2N2 Canada}

\author{J. H. Page}
\affiliation{Department of Physics and Astronomy, University of
Manitoba, Winnipeg, Manitoba, R3T 2N2 Canada}

\author{S.E. Skipetrov}
\affiliation{Universit\'{e} Joseph Fourier, Laboratoire de Physique et Mod\'{e}lisation des
Milieux Condens\'{e}s, CNRS, 25 Rue des Martyrs, BP 166, 38042 Grenoble, France}

\author{B.A. van Tiggelen}
\affiliation{Universit\'{e} Joseph Fourier, Laboratoire de Physique et Mod\'{e}lisation des
Milieux Condens\'{e}s, CNRS, 25 Rue des Martyrs, BP 166, 38042 Grenoble, France}

\date{\today}

%\begin{abstract}
%\end{abstract}

%\pacs{43.35.+d, 63.20.-e}

\maketitle

{\bf After exactly half a century of Anderson localization
\cite{anderson}, the subject is more alive than ever. Direct
observation of Anderson localization of electrons was always
hampered by interactions and finite temperatures. Yet, many
theoretical breakthroughs were made, highlighted by   finite-size
scaling \cite{g4}, the self-consistent theory \cite{vw} and
the numerical solution of the Anderson tight-binding model
\cite{Mack,irnum}. Theoretical understanding is based on
simplified models or approximations and comparison with experiment
is crucial. Despite a wealth of new experimental data, with
microwaves \cite{locamicro2,genacknature}, light \cite{
locaad,genack2,georg,georg2,segev}, ultrasound \cite{locasound}
and cold atoms \cite{atom,atom2,delande}, many questions remain,
especially for three dimensions. Here we report the first
observation of sound localization in a random three-dimensional
elastic network. We study the time-dependent transmission below
the mobility edge, and report ``transverse localization'' in three
dimensions, which has never been observed previously with any
wave. The data are well described by the self-consistent theory of
localization. The transmission reveals non-Gaussian statistics,
consistent with theoretical predictions. }

Most text books on condensed matter explain that the electronic
states in disordered conductors are extended plane or Bloch waves
with finite life times. This gives rise to ``Ohmic resistance",
 proportional to the length of the sample.
In the picture presented by Anderson \cite{anderson}, ``large''
disorder makes  electronic states localized in space. This offers
a mechanism to explain the widely observed metal-insulator
transitions \cite{MIT}. Scaling theory proposes a single
parameter, the Thouless conductance $g$, to describe the anomalous
length dependence of the resistance of a sample \cite{g4}. The
localized regime is characterized by the so-called Thouless
criterion $g<1$. While these ideas were first proposed for
electron localization, in the early eighties interest in classical
wave localization was raised \cite{john,pw3}, with the promise of
avoiding the difficulties caused by interactions in electronic
systems.  At the same time, the absence of bound states for
classical waves makes localization more challenging to achieve in
practice, with absorption as an additional concern.

\begin{figure}
\includegraphics[width=7.5cm]{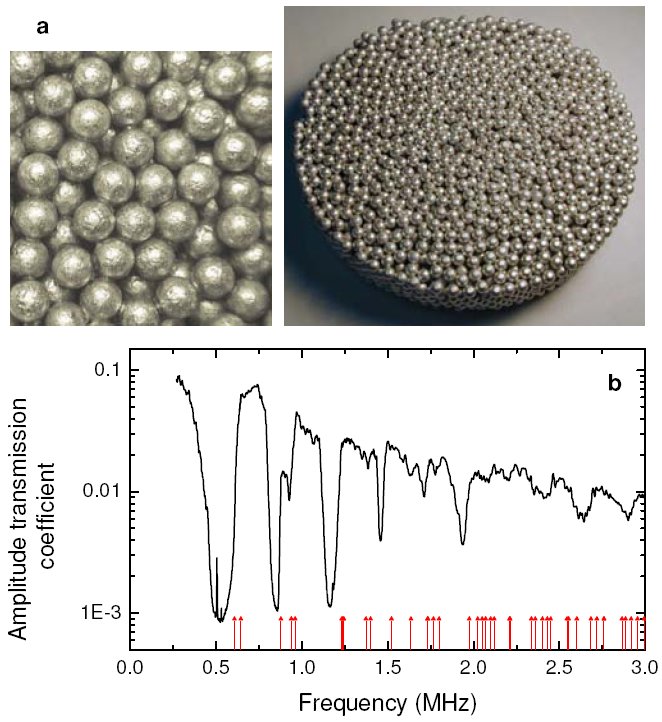}
\caption{ {\bf The random elastic network and corresponding
transmission coefficient. a}, Top view of one of the samples (right), along with a magnified image showing the mesoscale structure of the random elastic network (left).
%The picture on the left was taken before final cleaning, to minimize optical reflections and better reveal the local structure.
The
samples were made from $4.11 \pm 0.03$ mm diameter aluminum beads
brazed together at a volume fraction of approximately 55\%.
The  diameter (120 mm) of the slab-shaped samples
%The samples were slab shaped, as shown, with circular cross sections of width $W$ = 120 mm
was much larger than the thickness $L$, which ranged from 8.3
mm to 23.5 mm. {\bf b}, Frequency dependence of the amplitude
transmission coefficient for $L$ = 14.5 mm.  The arrows indicate
the resonance frequencies of isolated aluminum
beads.}\label{fig:sample}
\end{figure}

In this Letter we demonstrate Anderson localization of ultrasound
in a three-dimensional (3D) medium. Our samples are
single-component random networks made by brazing aluminum beads
together, see Fig.~\ref{fig:sample}a. With ultrasound we probe the
vibrational excitations of the network in the intermediate frequency regime (0.2 to
3 MHz), where the wavelength is comparable to the bead and pore
sizes. We use pulsed techniques to measure the amplitude
transmission coefficient, shown in Fig.~\ref{fig:sample}b. The
transmission spectrum exhibits a succession of band gaps and pass
bands, due to the overlapping resonances of the aluminium beads
\cite{turner}. Our study focuses on frequencies just below the
first band gap at 0.5 MHz, and at higher frequencies (1.6--3 MHz),
where it was possible to extract the phase and amplitude of the
coherent pulse for
longitudinal waves crossing the sample. Hence we were able to measure the longitudinal phase $v_p$ and group $v_g$ velocities, as well as the scattering mean free path $\ell$ \cite{pagesci,cowan}. Note that since absorption is negligible (see below), the attenuation of the coherent pulse gives $\ell$ directly.  Around 0.20 MHz,  $v_p= $ 1.75 km/s, $v_g=$ 2.1 km/s and $\ell=2.2$ mm, while around 2.4 MHz, $v_p=5.0$ km/s, $v_g=5.2$
km/s and $\ell= 0.6$ mm. This leads to a product of wave vector
$k$ and scattering mean free path $k\ell =1.6$ at 0.20 MHz and $k\ell=1.8$
at 2.4 MHz.  The values of $k\ell$ for shear waves, likely to
dominate, are not known but are probably roughly equal \cite{WRM}.
The small values of
$k\ell$ indicate that our samples are strongly scattering. For 3D
disorder localization is expected when $k\ell  \lesssim 1 $
\cite{houches}. Because the exact critical value is not known, the
ultrasonic waves at both frequencies are candidates to be Anderson
localized.

In previous reports on Anderson localization with classical waves,
absorption has been a major obstacle to reaching unambiguous
conclusions \cite{locaad,%localightgenack,
genacknature,georg,georg2,comments}. The following experiment is
capable of probing Anderson localization without being blurred by
absorption. We measure the spatially and time-resolved transmitted
intensity through our sample. Using  a quasi-point source that is
about a wavelength wide and a sub-wavelength-diameter detector
that scans the sample at various transverse positions $\rho$ in
the near field ($\rho = 0$ opposite to the source), the average
transmitted intensity $I(\rho, t)$ was determined \cite{Page1995}.
%The bandwidth was typically 5\% of the central frequency of the pulse.
From these measurements, we determine the ratio
$I(\rho,t)/I(0,t)$, probing the dynamic spreading of the intensity
in the transverse direction.  Any possible absorption factor
$\exp(-t/\tau_a)$ cancels in this ratio. For each $\rho$, a
transverse width $w_{\rho}(t)$ of  $I(\rho,t)$ can be defined by
setting $I(\rho,t)/I(0,t)\equiv \exp[-\rho^2/w_{\rho}^2(t)]$. If
the wave propagation  is diffuse, the spatial intensity profile is
Gaussian and $w_{\rho}^2(t) = 4Dt$ is independent of $\rho$. We
have observed this normal diffuse behavior in less strongly
scattering samples, and hence been able to demonstrate a way of
measuring the diffusion coefficient without the usual
complications due to boundary conditions and absorption
\cite{Page1995}.

\begin{figure}%[!h]
\includegraphics[width=7.5cm]{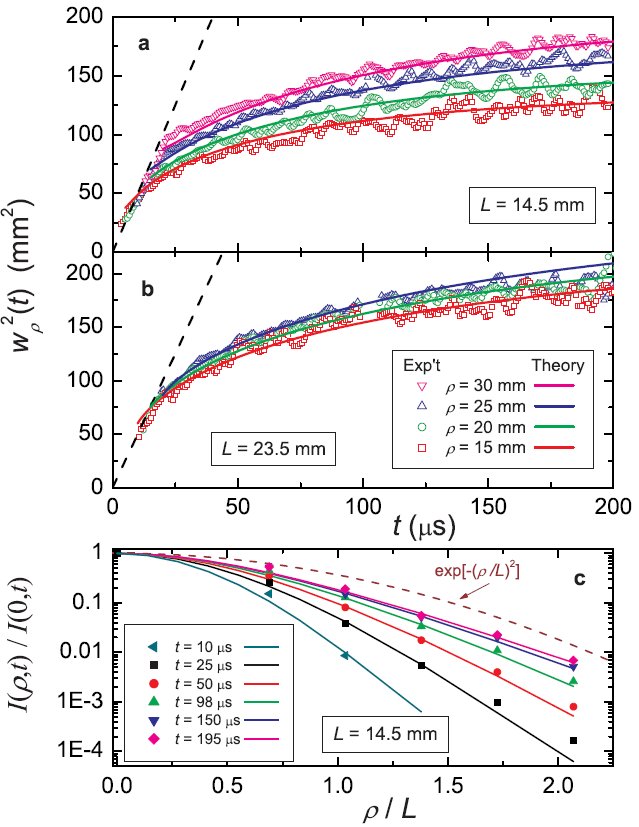}
\caption{ {\bf Transverse localization in three dimensions. a,b},
Temporal evolution of the effective width squared,
$w_{\rho}^2(t)$, of the transmitted intensity emanating from a
point source for several transverse displacements $\rho$ of the
detector and for two sample thicknesses. The frequency is 2.4 MHz. The solid curves are the best fits of our
self-consistent theory to the experimental data (symbols), from
which we determine $\ell^*_B = 2$ mm, $D_B = 11$
m$^2/$s, $\xi \approx 15$ mm for the $L=14.5$ mm sample and
$\xi \approx 7$ mm for the $L=23.5$ mm sample.  Other input parameters --- $k\ell = 1.8$ and the
internal reflection coefficient $R = 0.82$ --- were obtained from
independent measurements. The dashed line shows the linear
time-dependence of $w_{\rho}^2$ that would be seen for diffuse
waves, using $D=1.25$ m$^2/$s.  {\bf c}, Dependence of the
intensity ratio on distance $\rho$ at six different times, showing
the non-Gaussian profile that is found both experimentally
(symbols) and theoretically (solid curves). }
\label{fig:PointSource}
\end{figure}

For frequencies between 1.6 and 3 MHz, we observe dramatically different behavior, shown in Fig. \ref{fig:PointSource} for two representative samples at 2.4 MHz.
Instead of increasing linearly with propagation time,
$w_{\rho}^2(t)$ is seen to saturate. This saturation is
reminiscent of transverse localization, previously predicted in 3D
systems with 2D disorder \cite{pedro} and observed in 2D
disordered photonic crystals \cite{segev}. In our samples, the
disorder is clearly 3D since the scattering mean free path is much smaller
than the sample thicknesses ($10 < L/\ell < 40$). It is not at all clear
\emph{a priori } that transverse localization can occur with 3D
disorder. Scaling
theory \cite{g4} predicts a diffusivity $D(L)$ dependent on the sample size
$L$, so that  $w_{\rho}^2(t) = 4D(L)t$ still rises linearly with
time. The saturation of $w_{\rho}(t)$ could possibly be explained
by a  diffusivity $D(t)$ decaying with time
\cite{genack2,georg,georg2}. This would still lead to a Gaussian
transverse intensity profile and hence to $w_{\rho}^2$ constant
with $\rho$. However, this is not what we observe.

To describe the dynamics of the anomalous transverse confinement
of the multiply scattered waves, we apply  a novel version of the self-consistent (SC) theory of localization. The new element consists in
incorporating boundary conditions self-consistently \cite{Dz,Dz2}.
This theory provides a position dependent, dynamic diffusivity
kernel $D(\mathbf{r},t-t')$.  Near the mobility edge, the position
dependence of $D$ affects all aspects of wave transport
considerably. The SC theory requires as input the value for
$k\ell$,  the localization length $\xi$, the diffusion constant
$D_B$ free from macroscopic interferences, and the internal
reflection coefficient $R$.  In the model we replace the incident
focused beam by a point source at depth $\ell^*_B$. This is the
transport mean free path associated with diffusion in the absence
of macroscopic interferences, which ought to be negligible just
after the beam comes in. In Fig.\  \ref{fig:PointSource} we
compare the observed dynamics of transverse width to this theory.
Excellent agreement with the data is seen for \emph{all} $\rho$
with a \emph{single }set of parameters for each sample (solid curves), yielding,
in particular,  $\xi \approx 15$ mm and $\xi \approx 7$ mm for the thinner and thicker samples respectively. We attribute the smaller value of $\xi$ for the thicker sample to stronger scattering due to small differences in the microstructure. This is consistent with the strong sensitivity of $\xi$ to small changes in disorder that is expected near the localization transition. As $\rho$ increases,
the curves $w_{\rho}^2(t)$ move upwards, meaning that the observed
transverse profile $I(\rho, t)$ is not Gaussian.
Figure~\ref{fig:PointSource} shows that this behavior is well
captured by the theory, in which the position dependence of $D$ is
a crucial element. Any homogeneous absorption would not affect
Fig.~\ref{fig:PointSource}. We believe that this combination of
theory and experiment provides strong evidence for Anderson
localization of ultrasound near 2.4 MHz in our samples and allows
us to estimate the mobility edge $(k\ell)_c$, which we find to be approximately $1\%$ above the measured $k\ell = 1.8$.

\begin{figure}
\includegraphics[width=7.5cm]{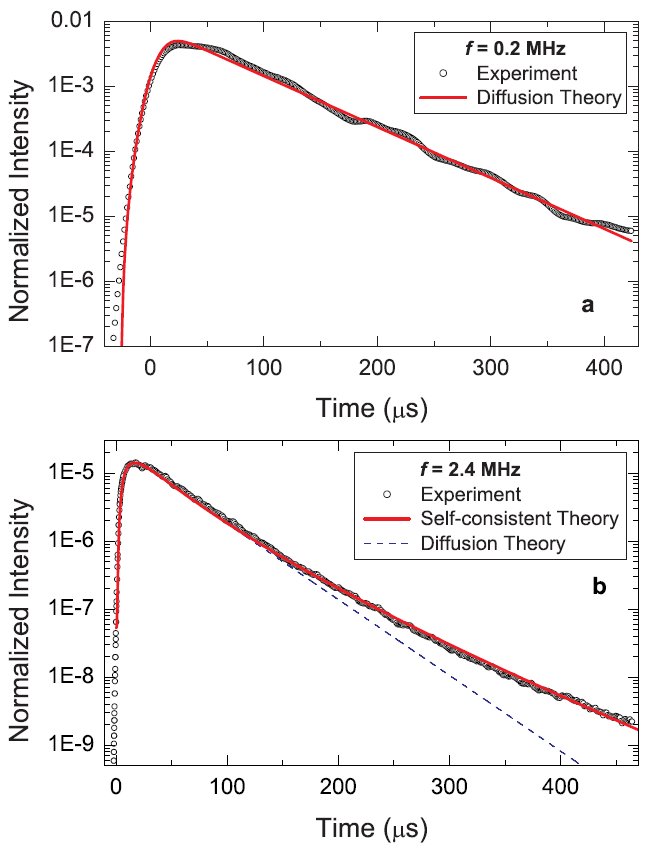}
\caption{ {\bf Averaged time-dependent transmitted intensity.}
Transmitted intensity, $I(t)$, normalized so that the peak of the
input pulse is unity and centered on $t = 0$, at representative
frequencies in the diffuse ({\bf a}) and localized ({\bf b})
regimes for the $L = 14.5$ mm sample investigated in Figs. 2a,c. In {\bf a}, the
fit to the diffusion theory with $R=0.85$ gives $D = 3.0$ m$^2$/s,
$\ell^* \simeq 2.5$ mm  and $\tau_a $ too large to be measurable.
In {\bf b}, the data are fitted by the SC theory (red curve), with
$\xi = 15$ mm, $\ell_B^* = 2$ mm, $D_B = 16$ m$^2/$s and $\tau_a
= 160 \, \mu$s.  For comparison, the dashed blue line shows the
long-time behaviour predicted by diffusion theory.}
\label{fig:I_PlaneWave}
\end{figure}

To find additional support for our conclusions, we have measured the
time-dependent transmission using an extended quasi plane-wave
source. Near 0.2 MHz, the average transmitted intensity $I(t)$ was
found to decay exponentially at long times
(Fig.~\ref{fig:I_PlaneWave}a), with the entire time dependence of
$I(t)$ being well described by diffusion theory\cite{Page1995}. We
conclude  that  multiply scattered ultrasound at $0.2$ MHz propagates in a normal, diffuse way. By contrast, for ultrasound propagating near 2.4 MHz, the time
dependence of $I(t)$ shows a quite different behaviour
(Fig.~\ref{fig:I_PlaneWave}b), with a non-exponential tail at long
times.  Similar behaviour has been reported before and is often
explained by a time-dependent reduction of the effective diffusion
coefficient $D(t)$ \cite{genack2,georg,georg2}. The solid curve in
Fig.~\ref{fig:I_PlaneWave}b is a fit to the SC theory, and gives
an excellent description of the experiment at all propagation
times. The good agreement between theory and experiment supports
our previous conclusion that ultrasound at 2.4 MHz is
localized. From the fits in Figs.~\ref{fig:PointSource} and
\ref{fig:I_PlaneWave}b and the relation $D_B = v_E \ell_B^*/3$ we
find transport velocities 3 to 5 times larger than the phase
velocity of longitudinal waves. Further theoretical work is needed
to understand these --- apparently large --- values for $v_E$ in
the localized regime.

\begin{figure*}[t]
\includegraphics[width=14cm]{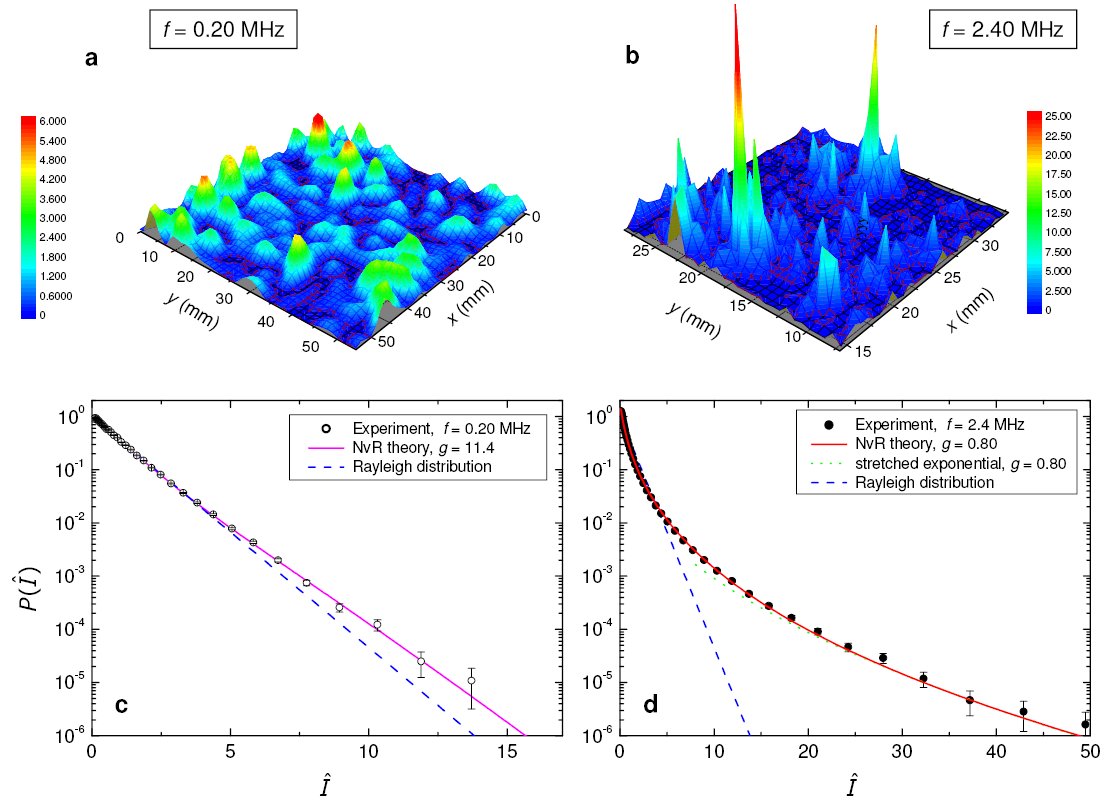}
\caption{ {\bf Statistical approach to localization.} Comparison
of the near field speckle patterns $I(x,y)/ \left<I\right>$ for
diffuse and localized waves observed at frequencies of 0.20 ({\bf
a}) and 2.4 MHz ({\bf b}). In {\bf a}, the speckle pattern is
typical for the diffuse regime, while panel {\bf b} reveals the
narrow intense spikes that we explain in terms of Anderson
localization. Note the different colour scales in the two figures.
{\bf c}, The measured probability distribution $P(\hat{I})$ of
normalized intensity ${\hat I} = I/\langle I \rangle$ at 0.2 MHz
(open circles) is close to the Rayleigh distribution (dashed blue
line). The solid magenta curve is best fit to the theory of
Ref.~\cite{theo} with $g = 11.4$. {\bf d}, At 2.4 MHz,
$P(\hat{I})$ (solid symbols) shows very large departures from the
Rayleigh distribution. The solid curve  shows the
theory~\cite{theo} with $g = 0.80$. At large $\hat{I} \gtrsim 25
$, the data can also be described by a stretched exponential
$P({\hat I}) \propto \exp( -2\sqrt{g {\hat I}}\; )$ with the same
value of $g$ (dotted curve). The large fluctuations $\langle
\hat{I}^2 \rangle = 2.74 $ and the large deviation from Rayleigh
statistics with $g < 1$ support our conclusion that Anderson
localization of sound has set in at frequencies near 2.4 MHz.}
\label{fig:stats}
\end{figure*}

We also address the statistical approach to Anderson
localization \cite{genacknature}. The normalized transmitted
intensity $\hat{I} \equiv I/ \left<I \right>$ was measured for a
large number of individual speckle spots in the near field, and
for a broad incident beam.  The intensity distributions $P(
\hat{I})$ for both frequencies are shown in Fig.~\ref{fig:stats}.
We have compared our data to the theory of Ref.~\cite{theo}, with
the Thouless conductance $g$ as the only free parameter in the
%least squares
fit. The agreement is excellent for $g=0.80 \pm 0.08
$ at 2.4 MHz and $g= 11.4 \pm 0.8$ at 0.2 MHz. The strong
deviation from Rayleigh statistics with $g<1$ observed at 2.4 MHz
is interpreted as a signature of localization \cite{genacknature}.
The remarkable agreement with theory, derived for the intensity in
the far field, and for $g\gg 1$,  reveals a universality of the
statistics of localized waves.

%In conclusion, we have used ultrasonic experiments, in conjunction
%with recent theoretical developments, to demonstrate Anderson
%localization of waves in a three-dimensional disordered system. We
%have discovered the phenomenon of 3D transverse localization,
%revealing how localization cuts off the dynamic transverse
%spreading of the intensity in 3D, with the width of the transverse
%intensity profile saturating at a value
%$w_{\rho}(t \rightarrow \infty)
% \sim$ 10--15 mm, which is of the order of the localization length $\xi$
%for our samples. Our findings demonstrate that ultrasonic
%experiments are very well suited for undertaking a complete study
%of one of the most fascinating concepts in condensed matter.

Our discovery of 3D transverse localization provides a powerful new approach for guiding future investigations of localization for any type of wave - not only for assessing whether or not the waves are localized but also for measuring the localization length.   By studying three different fundamental aspects of Anderson localization simultaneously, we have also shown the versatility of ultrasonic experiments, which are very well suited for undertaking a complete study of this phenomenon - one of the most fascinating in all condensed matter.  Our results suggest that it should now be feasible to determine critical exponents, by measuring the variation of the localization length with frequency near the mobility edge, as well as to investigate the sensitivity of the localization length to sample microstructure, the spatial correlations of localized waves in 3D, and the microscopic nature of localized states.  All of these experiments, either with sound or light, would make unique contributions to advancing our understanding of Anderson localization.

\vspace{.5cm}
 \textbf{METHODS}

 {\small The elastic network of aluminum beads was
created by precisely controlling the flux, alloy concentration and
temperature during brazing to form elastic bonds between the
beads while preserving their spherical shape.
The samples were made in the form of disc-shaped slabs, so as to minimize edge effects. After brazing, the samples were lightly polished to ensure that opposite faces of the slabs were flat and parallel, and carefully cleaned to remove any residue of the brazing process.
The samples were
waterproofed with very thin plastic walls to enable pulsed
ultrasonic immersion transducer techniques to be used
\cite{Page1995}, thereby ensuring that the samples remained dry
when immersed in a water tank between the generating and detecting
transducers. The coherent pulse, from which $v_p$, $v_g$, and
$\ell$ were determined, was then measured by scanning the position
of the sample in a plane parallel to the sample surface and
averaging the transmitted field over all positions \cite{pagesci,cowan}. These
measurements were made using large-diameter immersion transducers
to aid in the spatial averaging of the transmitted field and hence
in the extraction of the coherent component. The ratio of the
Fourier transforms of the transmitted and input signals gave the
amplitude transmission coefficient.

To measure $I(\rho, t)$, a quasi-point source was created by
focusing the pulsed ultrasonic beam onto a small aperture cut in
the tip of an acoustically-opaque cone-shaped screen.  The cone
shape was chosen so that edges of the beam could be effectively
blocked when the aperture was placed close to the sample being investigated, while at
the same time preventing significant stray sound being reflected
back towards the sample from the screen.  The transmitted field at
a transverse distance $\rho$ was measured using a miniature
hydrophone with a diameter of 400 $\mu$m, which is smaller than a wavelength, allowing the transmitted field to be detected in a
single speckle spot. The average transmitted intensity $I(\rho,
t)$ was determined for selected values of $\rho$ by scanning the
sample position in a grid over a very large number of independent
speckle spots \cite{Page1995}. To measure the time-dependent
transmission $I(t)$ for an extended quasi-plane-wave source, the
sample was placed deep in the far field of a small-diameter planar
transducer, and the transmitted speckle pattern was scanned by
moving the hydrophone with the sample fixed in position. For
measurements of both $I(\rho, t)$ and $I(t)$, the number of
speckle spots over which the intensity was averaged was typically
529--3025, and the bandwidth was set at $5\%$ of the central
frequency of the pulse by digitally filtering the transmitted
field before determining the dynamic intensity profiles.

The normalized intensity $\hat{I} \equiv I/ \left<I \right>$ at a
particular frequency was determined from the square of the
magnitude of the Fourier transform of the transmitted field in
each near-field speckle, normalized by the average intensity in
the speckle pattern. The results were then binned to determine $P(
\hat{I})$. To improve the statistics, the results were averaged
over up to 100 frequencies within a 5\% bandwidth where the
statistics were similar --- the same bandwidth as for the dynamic
measurements of $I(\rho, t)$ and $I(t)$.  In the localized regime,
where fluctuations are largest, the statistical accuracy of the
measured distribution was further improved by averaging over
different samples, which were found to exhibit the same
statistics. This allowed the error bars in Fig.~\ref{fig:stats}d
to be determined from the standard error in the mean.}

{\small \textbf{Acknowledgements}

Financial support from NSERC of Canada and a CNRS France-Canada
PICS project is gratefully acknowledged. The calculations in this
work have been performed on the cluster HealthPhy (CIMENT,
Grenoble).}

{\small \textbf{Competing Interests Statement} The authors declare
no competing financial interests.}

%\newpage

%\newpage

%\pagebreak[4]

\end{document}